\documentclass[reprint, notitlepage,superscriptaddress,preprintnumbers,nofootinbib,nobibnotes,amsmath,amssymb,aps,twocolumn]{revtex4-1}
\usepackage{graphicx}
\usepackage{dcolumn}
\usepackage{bm}
\usepackage{hyperref}
\usepackage{aas_macros,braket}

\begin{document}

\title{Parity violation in the CMB trispectrum from the scalar sector}

\author{Maresuke Shiraishi}
\affiliation{%
  Kavli Institute for the Physics and Mathematics of the Universe (Kavli IPMU, WPI), UTIAS, The University of Tokyo, Chiba, 277-8583, Japan
}

\date{\today}

\begin{abstract}
  Under the existence of chiral non-Gaussian sources during inflation, the trispectrum of primordial curvature perturbations can break parity. We examine signatures of the induced trispectrum of the cosmic microwave background (CMB) anisotropies. It is confirmed via a harmonic-space analysis that, as a consequence of parity violation, such a CMB trispectrum has nonvanishing signal in the $\ell_1 + \ell_2 + \ell_3 + \ell_4 = \text{odd}$ domain, which is prohibited in the concordance cosmology. When the curvature trispectrum is parametrized with Legendre polynomials, the CMB signal due to the Legendre dipolar term is enhanced at the squeezed configurations in $\ell$ space, yielding a high signal-to-noise ratio. A Fisher matrix computation results in a minimum detectable size of the dipolar coefficient in a cosmic-variance-limited-level temperature survey as $d_1^{\rm odd} = 640$. In an inflationary model where the inflaton field couples to the gauge field via an $f(\phi)(F^2 + F\tilde{F})$ interaction, the curvature trispectrum contains such a parity-odd dipolar term. We find that, in this model, the CMB trispectrum yields a high signal-to-noise ratio compared with the CMB power spectrum or bispectrum. Therefore, the $\ell_1 + \ell_2 + \ell_3 + \ell_4 = \text{odd}$ signal could be a promising observable of cosmological parity violation.
\end{abstract} 


\maketitle

\section{Introduction}

Large-scale anisotropy in the cosmic microwave background (CMB) radiation includes very clean information on the very early Universe. Their $N$-point correlators directly reflect statistical properties of primordial fluctuations. Parity of such correlators is a clue to identify the inflationary Lagrangian. It is widely known that the Chern-Simons-like interactions in gravity \cite{Lue:1998mq,Alexander:2004wk,Lyth:2005jf,Takahashi:2009wc,Alexander:2009tp,Satoh:2010ep,Soda:2011am,Shiraishi:2011st,Dyda:2012rj,Wang:2012fi,Zhu:2013fja} or the electromagnetic sector \cite{Sorbo:2011rz,Barnaby:2011vw,Barnaby:2012xt,Dimopoulos:2012av,Adshead:2013qp,Adshead:2013nka,Cook:2013xea,Caprini:2014mja,Ferreira:2014zia,Bartolo:2014hwa,Bielefeld:2014nza,Bartolo:2015dga,Namba:2015gja,Ferreira:2015omg,Obata:2016tmo,Maleknejad:2016qjz,Peloso:2016gqs} can create parity-violating signatures, not realized in Einstein gravity or the standard electromagnetism, in primordial metric perturbations. The most representative one is due to the chirality of the tensor mode. Asymmetry between two helicity modes $(\lambda = \pm 2)$ of gravitational waves (GWs) induces a nonvanishing parity-odd signal in the CMB temperature (T) and polarization (E/B) correlators. Under an assumption of isotropy and homogeneity of the Universe, nonzero TB and EB correlations are realized in the diagonal domain ($\ell_1 = \ell_2$) of harmonic space \cite{Lue:1998mq,Gluscevic:2010vv}.%
\footnote{ 
If the rotational invariance is broken, the distinctive parity-odd signal also appears in the off-diagonal domain of TT, TE, EE and BB \cite{Bartolo:2014hwa}.
} 
In inflationary models where source fields generate a strong non-Gaussian (NG) GW signal, the CMB tensor-mode bispectrum also becomes an informative observable. Parity-odd information of primordial GW NG is confined to the $\ell_1 + \ell_2 + \ell_3 =  \text{even (odd)}$ domain of TTB, TEB, EEB and BBB (TTT, TTE, TEE, TBB, EEE and EBB) \cite{Kamionkowski:2010rb,Shiraishi:2011st,Shiraishi:2012sn,Shiraishi:2013kxa,Namba:2015gja,Shiraishi:2016yun}. These vanish in the usual parity-conserving models and hence are distinctive observables of parity-violating ones.
No detection of such power spectra \cite{Saito:2007kt,Gerbino:2016mqb} and bispectra \cite{Shiraishi:2014roa,Shiraishi:2014ila,Ade:2015ava} indicates no significant breaking of parity symmetry in the tensor sector.%
\footnote{
See Refs.~\cite{Liu:2006uh,Feng:2006dp,Xia:2007qs,Cabella:2007br,Kahniashvili:2008va,Gubitosi:2009eu,Das:2009ys,Gruppuso:2011ci,Gluscevic:2012me,Gubitosi:2012rg,Kaufman:2013vbd,Kahniashvili:2014dfa,Gubitosi:2014cua,Galaverni:2014gca,Ade:2015cva,Ade:2015cao,Gruppuso:2015xza,Aghanim:2016fhp} for observational constraints on the TB or EB correlation sourced by late-time objects.}

In contrast, this paper focuses on parity-breaking effects on the scalar sector. Under a space inversion $ {\bf x} \to -{\bf x}$, the curvature perturbation is transformed according to $\zeta_{\bf k} \to \zeta_{-{\bf k}}$. A parity-invariant $N$-point correlator therefore obeys
\begin{eqnarray}
    \Braket{\prod_{n=1}^N \zeta_{{\bf k}_n}}
    = \Braket{\prod_{n=1}^N \zeta_{-{\bf k}_n}} ~. \label{eq:zeta4_Peven_cond}
\end{eqnarray}
With the reality condition, $\zeta_{\bf k} = \zeta_{- {\bf k}}^*$, Eq.~\eqref{eq:zeta4_Peven_cond} guarantees $\Braket{\prod_{n=1}^N \zeta_{{\bf k}_n}} \in \mathbb{R}$. In other words, if Eq.~\eqref{eq:zeta4_Peven_cond} is violated, the $N$-point correlator includes imaginary components.%
\footnote{ In a rotational-invariant case, such an imaginary signal is prohibited for $N \leq 3$ since rotational symmetry guarantees Eq.~\eqref{eq:zeta4_Peven_cond}; thus, the trispectrum ($ N = 4$) is the lowest-order correlator which can have an imaginary or equivalently parity-odd signal.}
Under a parity transformation $\zeta_{\bf k} \to \zeta_{-{\bf k}}$, harmonic coefficients of the induced CMB fluctuations are altered as $a_{\ell m} \to (-1)^\ell a_{\ell m}$ [see Eq.~\eqref{eq:alm_all}]. The parity-conserving condition~\eqref{eq:zeta4_Peven_cond} therefore leads to the confinement of a nonvanishing CMB signal to the $\sum_{n=1}^N \ell_n = \text{even}$ domain, and the breaking of Eq.~\eqref{eq:zeta4_Peven_cond} due to parity violation yields nonvanishing $\sum_{n=1}^N \ell_n = \text{odd}$ components.

In this paper we examine the CMB trispectrum due to the parity-odd curvature trispectrum respecting statistical homogeneity and rotational invariance, reading
\begin{eqnarray}
    \Braket{\prod_{n=1}^4 \zeta_{{\bf k}_n}}
 &=& (2\pi)^{3}
  \int d^3 K  
  \delta^{(3)}({\bf k}_1 + {\bf k}_2 + {\bf K}) \nonumber \\ 
&&  \delta^{(3)}({\bf k}_3 + {\bf k}_4 - {\bf K})
  t_{{\bf k}_3 {\bf k}_4}^{{\bf k}_1 {\bf k}_2}({\bf K}) \nonumber \\ 
&& 
  + (23~ {\rm perm})~, \label{eq:zeta4_ang_ave_def}
\end{eqnarray}
where  $t_{{\bf k}_3 {\bf k}_4}^{{\bf k}_1 {\bf k}_2}({\bf K}) = \left[  t_{-{\bf k}_3 -{\bf k}_4}^{-{\bf k}_1 -{\bf k}_2}(-{\bf K})  \right]^*$ holds. A parity-odd condition $\Braket{\prod_{n=1}^4 \zeta_{{\bf k}_n}} = - \Braket{\prod_{n=1}^4 \zeta_{-{\bf k}_n}}$ leads to  $t_{{\bf k}_3 {\bf k}_4}^{{\bf k}_1 {\bf k}_2}({\bf K}) = - t_{-{\bf k}_3 -{\bf k}_4}^{-{\bf k}_1 -{\bf k}_2}(-{\bf K})$, and hence $t_{{\bf k}_3 {\bf k}_4}^{{\bf k}_1 {\bf k}_2}({\bf K})$ becomes pure imaginary. In analogy with the analysis of a parity-even angle-dependent trispectrum \cite{Shiraishi:2013oqa}, let us consider a simple linear parametrization using the Legendre polynomials $P_n(x)$:
\begin{eqnarray}
  t_{{\bf k}_3 {\bf k}_4}^{{\bf k}_1 {\bf k}_2}({\bf K})
  &\equiv& i \sum_n 
  \left[ A_n P_n(\hat{k}_1 \cdot \hat{k}_3) + B_n P_n(\hat{k}_1 \cdot \hat{K})
     \right. \nonumber \\ 
&& \left.   + C_n P_n(\hat{k}_3 \cdot \hat{K})
     \right] \left[ ( \hat{k}_1 \times \hat{k}_3 ) \cdot \hat{K} \right]
  \nonumber \\ 
&& 
 P_\zeta(k_1)
 P_\zeta(k_3)
 P_\zeta(K) ~,
\end{eqnarray}
where $q \equiv |{\bf q}|$, $\hat{q} \equiv {\bf q} / q$ and ${\bf K} = -{\bf k}_1 - {\bf k}_2 = {\bf k}_3 + {\bf k}_4$. Imposing $t_{{\bf k}_3 {\bf k}_4}^{{\bf k}_1 {\bf k}_2}({\bf K}) = t^{{\bf k}_3 {\bf k}_4}_{{\bf k}_1 {\bf k}_2}(-{\bf K})$ [and hence $B_n = (-1)^n C_n$] and $A_n = B_n$ further simplifies this to
\begin{eqnarray}
t_{{\bf k}_3 {\bf k}_4}^{{\bf k}_1 {\bf k}_2}({\bf K})
&\equiv& i  \sum_n d_n^{\rm odd}
\left[ P_n(\hat{k}_1 \cdot \hat{k}_3) + P_n(\hat{k}_1 \cdot \hat{K})  \right. \nonumber \\ 
 &&\left. 
    + (-1)^n P_n(\hat{k}_3 \cdot \hat{K})
    \right] \left[ ( \hat{k}_1 \times \hat{k}_3 ) \cdot \hat{K} \right]
  \nonumber \\ 
 &&  P_\zeta(k_1) P_\zeta(k_3) P_\zeta(K) ~.  \label{eq:zeta4_dnodd_def}
\end{eqnarray}
This could be generated if there exist chiral NG sources during inflation. For example, in an inflationary model where the inflaton field $\phi$ couples to the U(1) gauge field via the coupling ${\cal L} = (1/4)f(\phi)(-F^2 + \gamma F\tilde{F})$ \cite{Caprini:2014mja,Bartolo:2015dga,Abolhasani:2015cve} [dubbed the $f(\phi)(F^2 + F\tilde{F})$ model henceforth], nonvanishing $d_{0}^{\rm odd}$ and $d_{1}^{\rm odd}$ are produced for $\gamma \neq 0$ (as we show in Appendix~\ref{appen:fFFtilde_zeta4}).

From Eq.~\eqref{eq:zeta4_dnodd_def}, we derive a CMB TTTT trispectrum by means of the all-sky and flat-sky formalisms. The flat-sky formula shows how the angular dependence in 3D ${\bf k}$ space in Eq.~\eqref{eq:zeta4_dnodd_def} is projected on 2D $\boldsymbol{\ell}$ space. From the all-sky formula, we confirm that the nonvanishing signal obeys $\ell_1 + \ell_2 + \ell_3 + \ell_4 = \text{odd}$ as expected. On the basis of the all-sky formula, we perform the Fisher matrix analysis for $d_{n}^{\rm odd}$. We then find that the prominent signal of the $d_1^{\rm odd}$ mode at the squeezed configurations ($k_1 \sim k_2 \gg K$ or $k_3 \sim k_4 \gg K$) enhances a signal-to-noise ratio of the TTTT trispectrum, and thus $d_1^{\rm odd} = 640$ is detectable in a cosmic-variance-limited-level (CVL-level) full-sky survey.

  In the $f(\phi)(F^2 + F\tilde{F})$ model, the sizes of the induced CMB power spectrum and bispectrum are determined by the ratio of the energy density of a vacuum expectation value (vev) of the electric component of the gauge field $\rho_E^{\rm vev} \equiv {\bf E}_{\rm vev}^2 /2$ to the inflaton energy density $\rho_\phi$. Reference~\cite{Bartolo:2015dga} showed this and found via a comparison between the theoretical CMB signal and the {\it Planck} 2015 constraints \cite{Ade:2015lrj,Ade:2015ava} that the upper bound on $ \rho_E^{\rm vev} / \rho_\phi$ is $10^{-9}$ ($4 \times 10^{-16}$) at $|\gamma| = 0$ ($1$) and becomes much tighter as $|\gamma|$ increases.%
  \footnote{For $|\gamma| > 1$, the bound from the bispectrum becomes more stringent than that from the power spectrum because of the difference of the scalings: $|\gamma|^6 / e^{8\pi|\gamma|}$ vs. $|\gamma|^3 / e^{4\pi|\gamma|}$ \cite{Bartolo:2015dga}.}
 In this paper we analyze, for the first time, the curvature trispectrum with $\gamma \neq 0$ and confirm that its amplitude also depends on $\rho_E^{\rm vev}/ \rho_\phi$. We find via Fisher matrix forecasts that a minimum detectable $\rho_E^{\rm vev} / \rho_\phi$ is $10^{-20}$ at $|\gamma| = 1$ and decreases as $|\gamma|$ grows by scaling like $|\gamma|^9 / e^{12\pi|\gamma|}$. Because of this, in the $|\gamma| > 1$ regime, the trispectrum is expected to yield a much more stringent bound on $\rho_E^{\rm vev} / \rho_\phi$ than the power spectrum and bispectrum. This indicates the superiority of $d_1^{\rm odd}$ as an observable of $\rho_E^{\rm vev} / \rho_\phi$.

The rest of the paper is organized as follows. In the next section we derive the CMB trispectrum generated from Eq.~\eqref{eq:zeta4_dnodd_def} by means of the flat-sky and the all-sky formalisms. In Sec.~\ref{sec:Fish} we find expected uncertainties on $d_n^{\rm odd}$ and $\rho_E^{\rm vev} / \rho_\phi$ in the $f(\phi)(F^2 + F\tilde{F})$ model through Fisher matrix computations. Section~\ref{sec:conclusion} is devoted to conclusions. Appendix~\ref{appen:fFFtilde_zeta4} presents the derivation of primordial curvature trispectrum realized in the $f(\phi)(F^2 + F\tilde{F})$ model.

\section{CMB trispectrum  from $d_n^{\rm odd}$}\label{sec:CMB}

We now study signatures of the parity-odd curvature trispectrum \eqref{eq:zeta4_dnodd_def} in the CMB trispectrum. First, we perform warm-up computations using the flat-sky approximation and learn the structure of the CMB parity-odd trispectrum schematically. After that, by means of the all-sky formalism, we derive a complete formula to use in Fisher matrix analysis (Sec.~\ref{sec:Fish}). 

Harmonic coefficients in the all-sky basis $a_{\ell m} = \int d^2 \hat{n} Y_{\ell m}^*(\hat{n}) \delta T(\hat{n})$ and the flat-sky basis $a_{\boldsymbol{\ell}} = \int d^2 \Theta e^{-i \boldsymbol{\ell} \cdot {\bf \Theta}} \delta T(\hat{n})$ are, respectively, given by \cite{Zaldarriaga:1996xe,Shiraishi:2010sm,Shiraishi:2010kd} 
\begin{eqnarray}
  a_{\ell m} &=& 
4\pi i^{\ell} \int \frac{d^3 k}{(2\pi)^{3}}
 \zeta_{\bf k} {\cal T}_{\ell}(k) Y_{\ell m}^*(\hat{k}) 
  ~, \label{eq:alm_all} \\
  a_{\boldsymbol{\ell}} &=&
  \int \frac{d^3 k}{(2\pi)^3} \zeta_{\bf k} \int_0^{\tau_0} d\tau  S_T(k,\tau) \nonumber \\ 
&& (2\pi)^2 \delta^{(2)}\left({\bf k}^\parallel D - \boldsymbol{\ell} \right) e^{i k_{z} D}  ~, \label{eq:alm_flat}
  \end{eqnarray}
where $\tau_0$ is the present conformal time, $S_T(k,\tau)$ is the scalar-mode source function for temperature fluctuations, $D \equiv \tau_0 - \tau$, ${\bf k}^\parallel \equiv (k_x, k_y)$ and ${\cal T}_\ell(k) \equiv \int_0^{\tau_0} d\tau S_T(k,\tau) j_\ell(kD)$.%
\footnote{In this paper we discuss the temperature trispectrum alone for simplicity, while the formulas derived here are straightforwardly extended to polarized trispectra by simply replacing ${\cal T}_\ell(k)$ or $S_T(k,\tau)$ with the E-mode polarization one \cite{Shiraishi:2010sm,Shiraishi:2010kd}.
}

\subsection{Flat-sky formalism}\label{subsec:flat}

The CMB trispectrum generated from the curvature trispectrum \eqref{eq:zeta4_ang_ave_def} reads
\begin{eqnarray}
  \Braket{  \prod_{n=1}^4 a_{\boldsymbol{\ell}_n} }
  &=& \left[ \prod_{n=1}^4 \int \frac{d^3 k_n}{(2\pi)^3}  \int_0^{\tau_0} d\tau_n  S_T(k_n,\tau_n) \right. \nonumber \\
&& \left.   (2\pi)^2 \delta^{(2)}({\bf k}_n^\parallel D_n - \boldsymbol{\ell}_n ) e^{i k_{nz} D_n} \right] \nonumber \\
  && (2\pi)^{3}
  \int d^3 K 
  \delta^{(3)}\left( {\bf k}_1 + {\bf k}_2 + {\bf K} \right) \nonumber \\
  &&
  \delta^{(3)}\left( {\bf k}_3 + {\bf k}_4 - {\bf K} \right)
 t_{{\bf k}_3 {\bf k}_4}^{{\bf k}_1 {\bf k}_2}({\bf K}) \nonumber \\ 
&&
 + (23~ {\rm perm})~, \label{eq:T4_flat_form}
\end{eqnarray}
where the wavevectors are projected onto the flat-sky space according to ${\bf k}_n = (\boldsymbol{\ell}_n / D_n, k_{nz})$ and ${\bf K} = (- \boldsymbol{\ell}_1 / D_1 - \boldsymbol{\ell}_2 / D_2, - k_{1z} - k_{2z}) = (\boldsymbol{\ell}_3 / D_3 + \boldsymbol{\ell}_4 / D_4, k_{3z} + k_{4z})$. The source function has a peak at around the recombination epoch $\tau_*$, and hence the signal for $D_1 \sim D_2 \sim D_3 \sim D_4 \sim r_* \equiv \tau_0 - \tau_*$ contributes dominantly to the $\tau$ integrals. Owing to this fact, we may approximate the delta function as 
\begin{eqnarray}
&& \delta^{(3)}({\bf k}_1 + {\bf k}_2 + {\bf K})
  \delta^{(3)}({\bf k}_3 + {\bf k}_4 - {\bf K}) \nonumber \\ 
  &&\quad \simeq
  r_*^4  
  \delta^{(2)}\left(\boldsymbol{\ell}_1 + \boldsymbol{\ell}_2 + {\bf J} \right)
  \delta^{(2)}\left(\boldsymbol{\ell}_3 + \boldsymbol{\ell}_4 - {\bf J}  \right) \nonumber \\ 
  &&\quad \int_{-\infty}^\infty \frac{dr}{2\pi} e^{i (k_{1z} + k_{2z} + K_z) r} \int_{-\infty}^\infty \frac{dr'}{2\pi} e^{i (k_{3z} + k_{4z} - K_z) r'},   
\end{eqnarray}
where we have introduced ${\bf J} \equiv {\bf K}^\parallel r_*$. Moreover, since we focus on the very high-$\ell$ region where the flat-sky approximation is reasonable enough, we can evaluate the $\tau$ integrals with $k_{nz} D_n / \ell_n \ll 1$. This simplifies the angle-dependent quantities to $\hat{k}_1 \cdot \hat{k}_3 \simeq \hat{\ell}_1 \cdot \hat{\ell}_3$, $\hat{k}_{1,3} \cdot \hat{K} \simeq \hat{\ell}_{1,3} \cdot \hat{J}$, and $(\hat{k}_1 \times \hat{k}_3) \cdot \hat{K} \simeq (k_{3z} r_* / \ell_3)  (\hat{J} \times \hat{\ell}_1) 
+ (k_{1z} r_* / \ell_1)  (\hat{\ell}_3 \times \hat{J})
+  (K_z r_* / J) (\hat{\ell}_1 \times \hat{\ell}_3)$, 
where $\hat{l} \times \hat{j} \equiv \hat{l}_{x} \hat{j}_{y} - \hat{l}_{y} \hat{j}_{x}$ and $\hat{l} \cdot \hat{j} \equiv \hat{l}_x \hat{j}_x + \hat{l}_y \hat{j}_y$. These reduce Eq.~\eqref{eq:T4_flat_form} to 
\begin{eqnarray}
\Braket{  \prod_{n=1}^4 a_{\boldsymbol{\ell}_n} }
  &\simeq& (2\pi)^2 \int d^2 J \delta^{(2)}\left(\boldsymbol{\ell}_1 + \boldsymbol{\ell}_2 + {\bf J} \right) \nonumber \\ 
&& \delta^{(2)}\left(\boldsymbol{\ell}_3 + \boldsymbol{\ell}_4 - {\bf J}  \right)
\sum_n d_n^{\rm odd} t_{\boldsymbol{\ell}_3 \boldsymbol{\ell}_4}^{\boldsymbol{\ell}_1 \boldsymbol{\ell}_2}({\bf J}, n) \nonumber \\
&& + (23~ {\rm perm}) ~,
\end{eqnarray}
where
\begin{eqnarray}
&&  t_{\boldsymbol{\ell}_3 \boldsymbol{\ell}_4}^{\boldsymbol{\ell}_1 \boldsymbol{\ell}_2}({\bf J}, n) 
 = \int_{-\infty}^\infty r_*^2 dr \int_{-\infty}^\infty r_*^2  dr' \nonumber \\ 
&&\, \left[ P_n(\hat{\ell}_1 \cdot \hat{\ell}_3)
    + P_n(\hat{\ell}_1 \cdot \hat{J})
    + (-1)^n P_n(\hat{\ell}_3 \cdot \hat{J})
    \right] \nonumber \\
&&\, \left[
       (\hat{J} \times \hat{\ell}_1)  
       {\cal B}_{\ell_1}(r) {\cal A}_{\ell_2}(r) \widetilde{\cal B}_{\ell_3}(r') {\cal A}_{\ell_4}(r') {\cal F}_J(r,r')   \right. \nonumber \\ 
 &&\, \left.
+ (\hat{\ell}_3 \times \hat{J}) 
\widetilde{\cal B}_{\ell_1}(r) {\cal A}_{\ell_2}(r)  {\cal B}_{\ell_3}(r') {\cal A}_{\ell_4}(r') {\cal F}_J(r,r')  \right. \nonumber \\ 
&&\, \left. +   
(\hat{\ell}_1 \times \hat{\ell}_3) {\cal B}_{\ell_1}(r) {\cal A}_{\ell_2}(r)
 {\cal B}_{\ell_3}(r') {\cal A}_{\ell_4}(r')
\widetilde{\cal F}_J(r,r')
    \right], \label{eq:T4_flat}
\end{eqnarray}
with
\begin{eqnarray}
  {\cal A}_\ell(r) &\equiv&
  \int_0^{\tau_0} d\tau 
  \int_{\ell / D}^{\infty} \frac{dk}{2\pi} \frac{1}{\sqrt{1 - \left(\frac{\ell}{kD}\right)^2  }} S_T(k,\tau) \nonumber \\ 
&&  \frac{2}{D^2} \cos\left[ k(r + D) \sqrt{1 - \left(\frac{\ell}{kD}\right)^2  } \right] 
  ~, \\
  {\cal B}_\ell(r) &\equiv& \int_0^{\tau_0} d\tau 
  \int_{\ell / D}^{\infty} \frac{dk}{2\pi} \frac{P_\zeta(k)}{\sqrt{1 - \left(\frac{\ell}{kD}\right)^2  }} S_T(k,\tau) \nonumber \\ 
&&
  \frac{2}{D^2} \cos\left[ k(r + D) \sqrt{1 - \left(\frac{\ell}{kD}\right)^2  } \right] 
  ~, \\
   {\cal F}_J(r,r')  &\equiv&  \int_{J / r_*}^\infty \frac{dK}{2\pi} \frac{P_\zeta(K)}{\sqrt{1 - \left(\frac{J}{Kr_*}\right)^2}} \nonumber \\ 
   && \frac{2}{r_*^2} \cos\left[ K(r -r') \sqrt{1 - \left(\frac{J}{Kr_*}\right)^2}  \right] ~,
\end{eqnarray}
and 
   \begin{eqnarray}
  \widetilde{\cal B}_\ell(r) &\equiv&
  - \int_0^{\tau_0} d\tau 
    \int_{\ell / D}^{\infty} \frac{dk}{2\pi}  \frac{k r_*}{\ell} P_\zeta(k) 
    S_T(k,\tau) \nonumber \\ 
&&
  \frac{2}{D^2} \sin\left[ k(r + D) \sqrt{1 - \left(\frac{\ell}{kD}\right)^2  } \right] 
  ~, \\
  \widetilde{\cal F}_J(r,r') &\equiv&
  -  \int_{J / r_*}^\infty \frac{dK}{2\pi} \frac{K r_*}{J} P_\zeta(K) \nonumber \\ 
&&
 \frac{2}{r_*^{2}} \sin\left[ K(r -r') \sqrt{1 - \left(\frac{J}{Kr_*}\right)^2}  \right] ~.
  \end{eqnarray}

\begin{figure}[t]
    \includegraphics[width=0.5\textwidth]{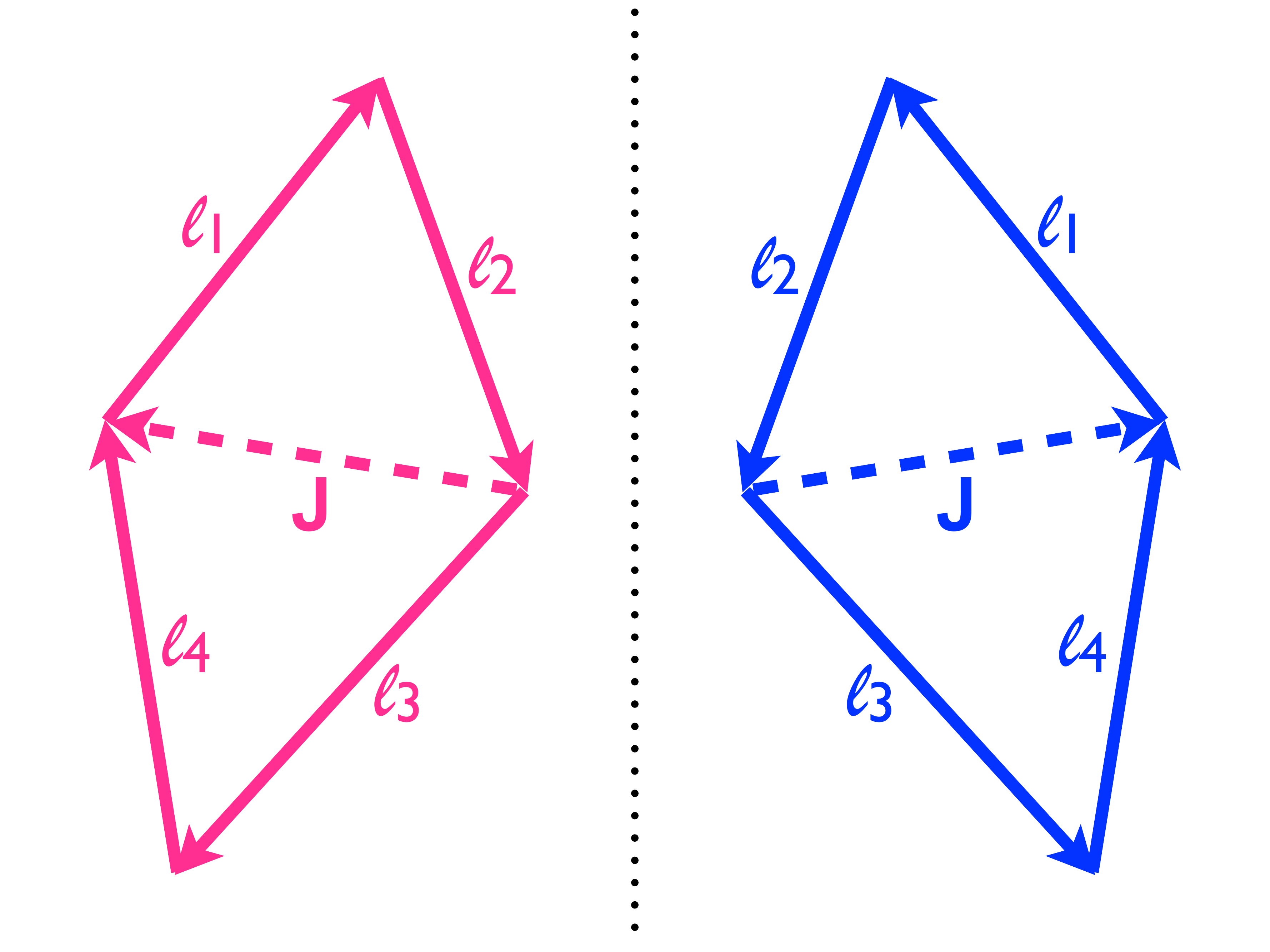}
    \caption{An $\boldsymbol{\ell}$-space configuration of the CMB trispectrum and its mirrored one. If these two trispectra have odd parity as in Eq.~\eqref{eq:T4_flat}, they take different signs with the same amplitude. A similar argument is established for the bispectrum case \cite{Kamionkowski:2010rb}.} \label{fig:box}
\end{figure}

The specific angular dependence between each ${\bf k}$ in Eq.~\eqref{eq:zeta4_dnodd_def} is projected onto CMB $\boldsymbol{\ell}$ space according to Eq.~\eqref{eq:T4_flat}. Owing to $P_n(\hat{\ell}_1 \cdot \hat{\ell}_3)$, $P_n(\hat{\ell}_{1} \cdot \hat{J})$ and $P_n(\hat{\ell}_{3} \cdot \hat{J})$, a variety of shapes are realized depending on $n$ \cite{Shiraishi:2013vja,Shiraishi:2013oqa}. The cross products $\hat{J} \times \hat{\ell}_{1}$, $\hat{\ell}_{3} \times \hat{J}$ and $\hat{\ell}_1 \times \hat{\ell}_{3}$ yield the breaking of mirror symmetry. Figure~\ref{fig:box} describes an example of mirrored images of $\boldsymbol{\ell}$-space configurations of two trispectra. As a consequence of parity-odd nature, these trispectra have opposite signs.

Under the absence of the angular dependence, Eq.~\eqref{eq:T4_flat} returns to the usual $\tau_{\rm NL}$-type trispectrum involving a peak at the squeezed limit $\ell_1 \simeq \ell_2 \gg J$ or $\ell_3 \simeq \ell_4 \gg J$ \cite{Hu:2001fa,Kogo:2006kh}. In the present case, however, due to the presence of the angular dependence, the squeezed-limit peak is modulated depending on $\boldsymbol{\ell}$-space configurations. For example, in the ``squeezed-collinear configurations'' (i.e., $\boldsymbol{\ell}_1 \parallel \boldsymbol{\ell}_3 \parallel {\bf J}$), it is highly suppressed because $\hat{J} \times \hat{\ell}_{1} \simeq \hat{\ell}_{3} \times \hat{J} \simeq \hat{\ell}_1 \times \hat{\ell}_{3} \simeq 0$. On the other hand, the modulation becomes mild in the ``squeezed-isosceles configurations'' (i.e., $\boldsymbol{\ell}_1 \perp {\bf J}$, $\boldsymbol{\ell}_3 \perp {\bf J}$ and $ \boldsymbol{\ell}_1 \parallel \boldsymbol{\ell}_3 $). The $n=1$ trispectrum $t_{\boldsymbol{\ell}_3 \boldsymbol{\ell}_4}^{\boldsymbol{\ell}_1 \boldsymbol{\ell}_2}({\bf J}, 1)$ then becomes comparable in size to the $\tau_{\rm NL}$-type trispectrum with $\tau_{\rm NL} = 1$. This seems to achieve high sensitivity to $d_1^{\rm odd}$ (see Sec.~\ref{sec:Fish}).

\subsection{All-sky formalism}

Using Eq.~\eqref{eq:alm_all}, the all-sky form is expressed as
\begin{eqnarray}
\Braket{\prod_{n=1}^4 a_{\ell_n m_n}} &=& 
  \left[ \prod_{n=1}^4 4\pi i^{\ell_n} \int \frac{d^3 k_n}{(2\pi)^{3}}
   {\cal T}_{\ell_n}(k_n) Y_{\ell_n m_n}^*(\hat{k}_n)  \right] \nonumber \\
  && (2\pi)^{3}
  \int d^3 K  
  \delta^{(3)}({\bf k}_1 + {\bf k}_2 + {\bf K}) \nonumber \\ 
&&  \delta^{(3)}({\bf k}_3 + {\bf k}_4 - {\bf K})
  t_{{\bf k}_3 {\bf k}_4}^{{\bf k}_1 {\bf k}_2}({\bf K})\nonumber \\ 
  && + (23~ {\rm perm})~. \label{eq:T4_all_form}
\end{eqnarray}
This is computed following the same procedure as Refs.~\cite{Shiraishi:2010kd,Shiraishi:2011st,Shiraishi:2013oqa}. We start by expressing the angle-dependent quantities with spherical harmonics. Using the identities
\begin{eqnarray}
  P_L(\hat{q}_1 \cdot \hat{q}_2) &=& \frac{4\pi}{2L+1} \sum_{M} Y_{LM}^*(\hat{q}_1) Y_{LM}(\hat{q}_2) ~, \\
i \, ( \hat{q}_1 \times \hat{q}_2 ) \cdot \hat{q}_3 
 &=& \sqrt{6} \left( \frac{4\pi}{3}  \right)^{3/2} \sum_{M_1 M_2 M_3}  \nonumber \\
&& Y_{1 M_1} (\hat{q}_1) Y_{1 M_2} (\hat{q}_2)  Y_{1 M_3} (\hat{q}_3) \nonumber \\
&&
\left(
  \begin{array}{ccc}
  1 & 1 & 1 \\
  M_1 & M_2 & M_3
  \end{array}
  \right) ~,
\end{eqnarray}
and the law of the addition of spherical harmonics, we rewrite the parity-odd trispectrum \eqref{eq:zeta4_dnodd_def} as
\begin{eqnarray}
  t_{{\bf k}_3 {\bf k}_4}^{{\bf k}_1 {\bf k}_2}({\bf K})
  &=& P_\zeta(k_1)
 P_\zeta(k_3)
 P_\zeta(K) 
 \sum_n d_n^{\rm odd} \sum_{L_1' L_3' L_K }
  \nonumber \\ 
&&  G_{L_1' L_3' L_K}^n
 \sum_{M_1' M_3' M_K} Y_{L_1' M_1'}^*(\hat{k}_1) Y_{L_3' M_3'}^*(\hat{k}_3)
  \nonumber \\ 
&&
 Y_{L_K M_K}^* (\hat{K})
 \left(
  \begin{array}{ccc}
  L_1' & L_3' & L_K\\
  M_1' & M_3' & M_K
  \end{array}
  \right) ~,
\end{eqnarray}
where
\begin{eqnarray}
  G_{L_1' L_3' L_K}^n &\equiv&
\frac{4\pi}{2n + 1}
 \sqrt{6} \left( \frac{4\pi}{3}  \right)^{3/2} \nonumber \\ 
&&  \left[(-1)^{n}
   h_{n 1 L_1'} 
   h_{n 1 L_3'}
   \delta_{L_K, 1}
    \left\{
  \begin{array}{ccc}
  L_1' & L_3' & L_K \\
  1 & 1 & n
  \end{array}
  \right\}
  \right. \nonumber \\ 
  && \left. 
  +  (-1)^{n}   h_{n 1 L_1'}\delta_{L_3', 1} h_{n 1 L_K} 
\left\{
  \begin{array}{ccc}
  L_1' & L_K & L_3'\\
  1 & 1 & n
  \end{array}
  \right\}  
    \right. \nonumber \\
&& \left.
    + 
    \delta_{L_1', 1}
   h_{n 1 L_3'}
   h_{n 1 L_K} 
\left\{
  \begin{array}{ccc}
  L_3' & L_K & L_1'\\
  1 & 1 & n
  \end{array}
  \right\}    
  \right] ~,
\end{eqnarray}
with $h_{l_1 l_2 l_3} \equiv \sqrt{\frac{(2 l_1 + 1)(2 l_2 + 1)(2 l_3 + 1)}{4 \pi}} \left(
  \begin{array}{ccc}
  l_1 & l_2 & l_3 \\
   0 & 0 & 0 
  \end{array}
  \right)$. The selection rules of $h_{l_1 l_2 l_3}$ and the Kronecker delta restrict $L_1'$, $L_3'$ and $L_K$ to $|n \pm 1|$ or $1$. The delta functions are also decomposed according to 
\begin{eqnarray}
&& \delta^{(3)}\left( \sum_{n=1}^3 {\bf q}_n \right) 
= 8 \int_0^\infty r^2 dr \nonumber \\ 
&&\qquad
\left[ \prod_{n=1}^3 \sum_{L_n M_n} 
 j_{L_n}(q_n r) 
Y_{L_n M_n}^*(\hat{q}_n) \right] 
\nonumber \\
&&\qquad (-1)^{\frac{L_1 + L_2 + L_3}{2}}
h_{L_1 L_2 L_3} 
\left(
  \begin{array}{ccc}
  L_1 & L_2 & L_3 \\
  M_1 & M_2 & M_3 
  \end{array}
  \right) ~.
\end{eqnarray}
We next perform the angular integrals of the products of spherical harmonics employing the identities 
\begin{eqnarray}
  \int d^2 \hat{q} \prod_{n=1}^2 Y_{L_n M_n}(\hat{q})
  &=& (-1)^{M_1} \delta_{L_1, L_2} \delta_{M_1, -M_2} ~, \\
  \int d^2 \hat{q} \prod_{n=1}^3 Y_{L_n M_n}(\hat{q}) &=& h_{L_1 L_2 L_3}
\left(
  \begin{array}{ccc}
  L_1 & L_2 & L_3 \\
  M_1 & M_2 & M_3 
  \end{array}
  \right).
  \end{eqnarray}
Finally, summing over the angular momenta in the resultant Wigner symbols, we obtain 
\begin{eqnarray}
  \Braket{\prod_{n=1}^4  a_{\ell_n m_n}
  }
  &=&   
  \sum_{J\mu} (-1)^{\mu}
  \left(
\begin{array}{ccc}
 \ell_1 & \ell_2 & J \\
 m_1 & m_2 & -\mu
\end{array}
\right) \nonumber \\
&& \left(
\begin{array}{ccc}
 \ell_3 & \ell_4 & J\\
 m_3 & m_4 & \mu
\end{array}
\right) \sum_n d_n^{\rm odd} t_{\ell_3 \ell_4 
}^{\ell_1 \ell_2
}(J,n) \nonumber \\ 
&& + (23~ {\rm perm}) ~,
  \end{eqnarray}
where
\begin{eqnarray}
  t_{\ell_3 \ell_4}^{\ell_1 \ell_2}(J,n)
  &=& i
\sum_{L_1 L} \sum_{L_3 L'} O_{\ell_3 \ell_4; L_3 L'}^{\ell_1 \ell_2; L_1 L}(J,n) 
    \nonumber \\  
&& 
  \int_0^\infty r^2 dr \int_0^\infty r'^2 dr' 
  \beta_{\ell_1 L_1}
(r)
\alpha_{\ell_2}
(r)
\nonumber \\
&&
\beta_{\ell_3 L_3}
(r')
\alpha_{\ell_4}
(r')
F_{L L'}(r,r') ~, \label{eq:T4_all}
  \end{eqnarray}
with
\begin{eqnarray}
  O_{\ell_3 \ell_4; L_3 L'}^{\ell_1 \ell_2; L_1 L}(J,n)
  &\equiv& (2J+1)
  (-1)^{\frac{ \ell_1 + \ell_3  + L_1  + L_3 + L + L' + 1}{2}} \nonumber \\ 
 && 
 h_{L_1 \ell_2 L} h_{L_3 \ell_4 L'} 
 \sum_{L_1' L_3' L_K}
(-1)^{L_3'} G_{L_1' L_3' L_K}^n \nonumber \\
 && h_{\ell_1 L_1 L_1'} h_{\ell_3 L_3 L_3'}  h_{L L' L_K} 
    \left\{
\begin{array}{ccc}
 J & L & L_1'  \\
 L_1 & \ell_1 & \ell_2 
\end{array}
\right\} \nonumber \\
 &&
    \left\{
\begin{array}{ccc}
 J & L' & L_3'  \\
 L_3 & \ell_3 & \ell_4 
\end{array}
\right\}
\left\{
\begin{array}{ccc}
 L_1' & L_3' & L_K  \\
 L' & L & J 
\end{array}
\right\} \label{eq:Ogeo} 
\end{eqnarray}
and 
\begin{eqnarray}
  \alpha_\ell
  (r) &\equiv& \frac{2}{\pi} \int_0^\infty k^2 dk {\cal T}_\ell
  (k)  j_\ell(k r) ~, \\
  \beta_{\ell L}(r) &\equiv& \frac{2}{\pi} \int_0^\infty k^2 dk P_\zeta(k) {\cal T}_\ell(k) j_{L}(k r)  ~, \\
      F_{L L'}(r,r') &\equiv& \frac{2}{\pi} \int_0^\infty K^2 dK P_\zeta(K) j_{L}(K r)  j_{L'}(K r') 
  ~.
\end{eqnarray} 
The summation ranges of $\sum_{L_1' L_3' L_K}$, $\sum_{L_1 L}$ and $\sum_{L_3 L'}$ are determined by $\ell_1$, $\ell_2$, $\ell_3$, $\ell_4$, $J$ and $n$. The nonzero signal of $t_{\ell_3 \ell_4}^{\ell_1 \ell_2}(J,n)$ is confined to $|\ell_1 - \ell_2| \leq J \leq \ell_1 + \ell_2$ and $|\ell_3 - \ell_4| \leq J \leq \ell_3 + \ell_4$. Moreover, due to the even $l_1 + l_2 + l_3$ filtering by $h_{l_1 l_2 l_3}$ and the odd $L_1' + L_3' + L_K$ filtering by $G_{L_1' L_3' L_K}^n$, nonvanishing $t_{\ell_3 \ell_4}^{\ell_1 \ell_2}(J,n)$ obeys $\ell_1 + \ell_2 + \ell_3 + \ell_4 = \text{odd}$ and takes pure imaginary numbers. This is an expected property of the parity-odd trispectrum. Note that $t_{\ell_3 \ell_4}^{\ell_1 \ell_2}(J,n) = t^{\ell_3 \ell_4}_{\ell_1 \ell_2}(J,n)$ holds.

In the Fisher matrix analysis in the next section, we estimate the trispectrum with $P_\zeta(k) = 2\pi^2 A_S k^{-3}$. We then have a useful analytic formula as 
\begin{eqnarray}
  F_{LL'}(r_*, r_*) &=&   
\frac{\pi^2 }{2} A_S \frac{ \Gamma(\frac{L+L'}{2})}
{\Gamma(\frac{L-L'+3}{2}) \Gamma(\frac{L'-L+3}{2})
  \Gamma(\frac{L+L'+4}{2}) } \nonumber  \\
&&\qquad\qquad\qquad\qquad (\text{for }L+L' > 0)~.
\end{eqnarray}
In the Sachs-Wolfe (SW) limit, i.e., ${\cal T}_\ell(k) \to - j_\ell(kr_*) / 5$, the $\alpha$ and $\beta$ functions become $\alpha_\ell(r) \to - \delta(r-r_*) / (5r_*^2)$ and $\beta_{\ell L}(r_*) \to - F_{\ell L}(r_*, r_*) / 5$, respectively; thus, the trispectrum can be reduced to
\begin{eqnarray}
  t_{\ell_3 \ell_4}^{\ell_1 \ell_2}(J,n)
  &\to&  i
\sum_{L_1 L} \sum_{L_3 L'} O_{\ell_3 \ell_4; L_3 L'}^{\ell_1 \ell_2; L_1 L}(J,n) 
5^{-4} F_{\ell_1 L_1}(r_*,r_*) \nonumber \\
&& F_{\ell_3 L_3}(r_*,r_*)
F_{L L'}(r_*,r_*) ~. \label{eq:T4_all_SW}
  \end{eqnarray}
This should be reasonable for small $\ell$.

As shown in the next section, the signal-to-noise ratio of the $d_1^{\rm odd}$ trispectrum converges with very few $J$'s. For such a case, we may use the following approximation. Since $\alpha_\ell(r)$ is sharply peaked at $r \sim r_*$, the $r$ and $r'$ integrals in Eq.~\eqref{eq:T4_all} are determined by the signal at $r \simeq r' \simeq r_*$. When $J$ and $n$ are small, $L$ and $L'$ are as small as they are (due to the selection rules of the Wigner symbols), and hence $F_{LL'}(r, r')$ varies very slowly at $r \sim r' \sim r_*$. These facts lead to 
\begin{eqnarray}
  t_{\ell_3 \ell_4}^{\ell_1 \ell_2}(J,n)
  &\approx& i \sum_{L_1 L} \sum_{L_3 L'} O_{\ell_3 \ell_4; L_3 L'}^{\ell_1 \ell_2; L_1 L}(J,n) 
    \nonumber \\  
&& 
R_{\ell_1 L_1 \ell_2}
R_{\ell_3 L_3 \ell_4}
F_{L L'}(r_*,r_*) ~, \label{eq:T4_all_smallJ}
  \end{eqnarray}
where 
\begin{eqnarray}
R_{\ell_1 L_1 \ell_2} \equiv \int_0^\infty r^2 dr 
\beta_{\ell_1 L_1}(r) 
\alpha_{\ell_2}(r) ~.  
\end{eqnarray}
This kind of approximation has been utilized in the analysis of the $\tau_{\rm NL}$-type trispectrum \cite{Pearson:2012ba} or the quadrupolar angle-dependent trispectrum \cite{Shiraishi:2013oqa}.

The approximations \eqref{eq:T4_all_SW} and \eqref{eq:T4_all_smallJ} make the Fisher matrix computations (shown in the next section) feasible.

\section{Fisher matrix forecasts} \label{sec:Fish}

Let us consider the CMB trispectrum measurement employing the angle-averaged quantity, defined as 
\begin{eqnarray}
  T_{\ell_3 \ell_4}^{\ell_1 \ell_2}(J) &\equiv&
  (2J + 1) 
  \sum_{m_1 m_2 m_3 m_4 \mu} (-1)^{\mu}
\left(
  \begin{array}{ccc}
  \ell_1 & \ell_2 & J \\
  m_1 & m_2 & -\mu 
  \end{array}
  \right)
  \nonumber \\ 
&& 
\left(
  \begin{array}{ccc}
  \ell_3 & \ell_4 & J \\
  m_3 & m_4 & \mu 
  \end{array}
  \right)
  \Braket{\prod_{n=1}^4 a_{\ell_n m_n}} ~. \label{eq:T4_ang_ave_def}
\end{eqnarray}
Using a diagonal covariance matrix approximation, the Fisher matrix for the amplitude parameter $A_{\rm tris} \, [\propto T_{\ell_3 \ell_4}^{\ell_1 \ell_2}(J)]$ is reduced to   
\begin{eqnarray}
  F_{A_{\rm tris}} = \sum_{\ell_1 \ell_2 \ell_3 \ell_4 J} \frac{\left|\hat{T}_{\ell_3 \ell_4}^{\ell_1 \ell_2}(J) \right|^2 }{24 (2J+1) \prod_{n=1}^4 C_{\ell_n}} 
  ~, \label{eq:Fisher}
  \end{eqnarray}
where $\hat{T}_{\ell_3 \ell_4}^{\ell_1 \ell_2}(J) = T_{\ell_3 \ell_4}^{\ell_1 \ell_2}(J) / A_{\rm tris}$ and  $C_\ell$ is the CMB power spectrum. The reduced trispectrum $t_{\ell_3 \ell_4 }^{\ell_1 \ell_2}(J)$, defined in
\begin{eqnarray}
  \Braket{\prod_{n=1}^4  a_{\ell_n m_n}}
  &=&   
  \sum_{J\mu} (-1)^{\mu}
  \left(
\begin{array}{ccc}
 \ell_1 & \ell_2 & J \\
 m_1 & m_2 & -\mu
\end{array}
\right) \nonumber \\ 
&& \left(
\begin{array}{ccc}
 \ell_3 & \ell_4 & J\\
 m_3 & m_4 & \mu
\end{array}
\right) t_{\ell_3 \ell_4 }^{\ell_1 \ell_2}(J)
\nonumber \\ 
&&
+ (23~ {\rm perm}) ~,
\end{eqnarray}
is related to $T_{\ell_3 \ell_4}^{\ell_1 \ell_2}(J)$ according to
\begin{eqnarray}
T_{\ell_3 \ell_4}^{\ell_1 \ell_2}(J) 
&=& P_{\ell_3 \ell_4}^{\ell_1 \ell_2}(J) + P^{\ell_3 \ell_4}_{\ell_1 \ell_2}(J) \nonumber \\
&&+ 
(2J+1) \sum_{J'} 
 (-1)^{\ell_2 + \ell_3} 
 \left\{
  \begin{array}{ccc}
  \ell_1 & \ell_2 & J \\
  \ell_4 & \ell_3 & J' 
  \end{array}
 \right\} \nonumber \\ 
&& \quad \left[ P_{\ell_2 \ell_4}^{\ell_1 \ell_3}(J') + P^{\ell_2 \ell_4}_{\ell_1 \ell_3}(J') \right]
\nonumber \\ 
&& 
+ (2J+1) \sum_{J'} (-1)^{J+J'} 
\left\{
  \begin{array}{ccc}
  \ell_1 & \ell_2 & J \\
  \ell_3 & \ell_4 & J' 
  \end{array}
 \right\} \nonumber \\ 
&& \quad \left[ P_{\ell_3 \ell_2}^{\ell_1 \ell_4}(J') + P^{\ell_3 \ell_2}_{\ell_1 \ell_4}(J') \right] ~,
\end{eqnarray} 
where
\begin{eqnarray}
  P_{\ell_3 \ell_4}^{\ell_1 \ell_2}(J) &=& t_{\ell_3 \ell_4}^{\ell_1 \ell_2}(J)
  + (-1)^{\ell_1 + \ell_2 + J} t_{\ell_3 \ell_4}^{\ell_2 \ell_1}(J) \nonumber \\ 
 &&
  + (-1)^{\ell_3 + \ell_4 + J} t_{\ell_4 \ell_3}^{\ell_1 \ell_2}(J) \nonumber \\ 
 &&
  + (-1)^{\ell_1 + \ell_2 + \ell_3 + \ell_4} t_{\ell_4 \ell_3}^{\ell_2 \ell_1}(J) ~.
\end{eqnarray}
The expected $1\sigma$ uncertainty on $A_{\rm tris}$ is given by $\Delta A_{\rm tris} = 1 / \sqrt{F_{A_{\rm tris}}}$.

In this section we perform the sensitivity analysis on the CMB trispectrum by computing Eq.~\eqref{eq:Fisher}. We then consider a full-sky CVL-level survey of temperature anisotropies up to $\ell = 2000$, and therefore Eq.~\eqref{eq:Fisher} does not include any instrumental uncertainties.

\subsection{Expected uncertainties on $d_n^{\rm odd}$}

We first investigate the sensitivity to $d_n^{\rm odd}$. We then focus on the lowest two modes, i.e., $n=0$ and $1$, which are produced in the $f(\phi)(F^2 + F\tilde{F})$ model as shown below.

\begin{figure}[t]
    \includegraphics[width=0.5\textwidth]{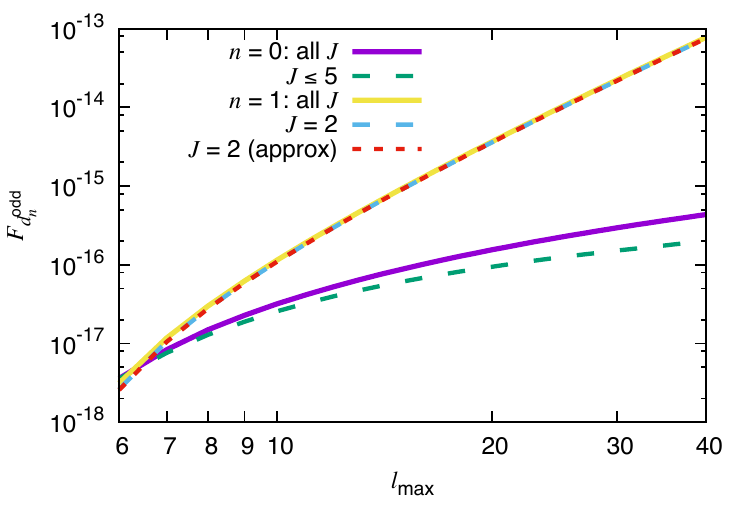}
    \caption{Fisher matrices for $d_0^{\rm odd}$ and $d_1^{\rm odd}$ in the SW limit as a function of $\ell_{\rm max}$. The two solid lines (purple and yellow) show the results obtained by summing over all possible $J$'s (satisfying $|\ell_1 - \ell_2| \leq J \leq \ell_1 + \ell_2$ and $|\ell_3 - \ell_4| \leq J \leq \ell_3 + \ell_4$), while the other lines are computed with a few $J$'s: $J \leq 5$ (green) and $J = 2$ (cyan and red). The red dotted line is obtained using the approximation~\eqref{eq:T4_P_app}.} \label{fig:Fish_SW_d0_d1}
\end{figure}

Figure~\ref{fig:Fish_SW_d0_d1} shows the Fisher matrices for $d_0^{\rm odd}$ and $d_1^{\rm odd}$ computed with the SW formula~\eqref{eq:T4_all_SW}. The solid ``all $J$'' lines correspond to the results obtained from all possible $J$'s, while the dashed or dotted lines are estimated with only a couple of $J$'s. It is easy to confirm from this figure that, for the $n = 1$ case, the result from $J = 2$ alone (cyan dashed line) completely overlaps that from all $J$'s (yellow solid line). Such rapid convergence of $F_{d_1^{\rm odd}}$ is realized by virtue of the enhanced signal at the squeezed configurations. In contrast, for the $n=0$ case, the Fisher matrix does not converge regardless of adding up to $J = 5$ (compare green dashed line with purple solid one), indicating the absence of the squeezed-limit enhancement. For this reason, a significant signal-to-noise ratio is not expected; thus, we do not analyze $d_0^{\rm odd}$ any further. The red dotted line represents $F_{d_1^{\rm odd}}$ from $J=2$ alone, obtained by employing the following approximation:
\begin{eqnarray}
T_{\ell_3 \ell_4}^{\ell_1 \ell_2}(J) 
\approx P_{\ell_3 \ell_4}^{\ell_1 \ell_2}(J) + P^{\ell_3 \ell_4}_{\ell_1 \ell_2}(J)~. \label{eq:T4_P_app}
\end{eqnarray}
This is reasonable in the case where the signal in the squeezed configurations dominates the  Fisher matrix \cite{Hu:2000ee}. As shown in Fig.~\ref{fig:Fish_SW_d0_d1}, in the $n=1$ case, this reproduces the exact result with an uncertainty of $\lesssim 1\%$. This achieves a significant reduction in computational complexity, making estimations with $\ell_{\rm max} = 2000$ feasible.

\begin{figure}[t]
    \includegraphics[width=0.5\textwidth]{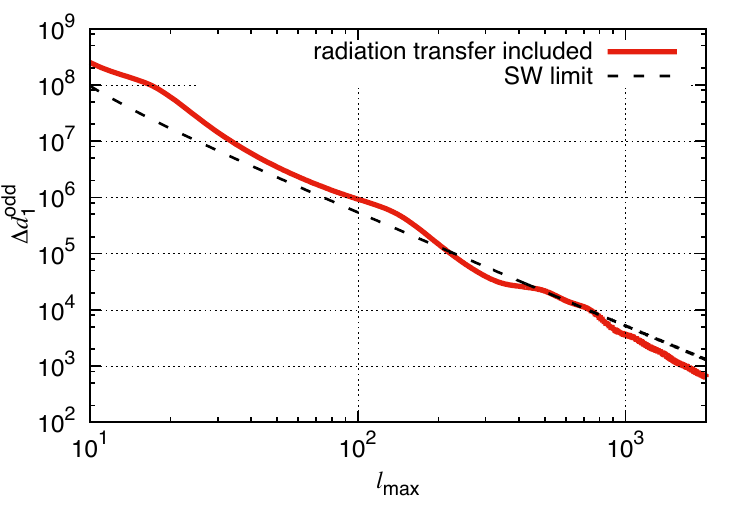}
    \caption{Expected $1\sigma$ errors on $d_1^{\rm odd}$ obtained through the Fisher matrix estimations with only the $J = 2$ signal and Eq.~\eqref{eq:T4_P_app}, justified in Fig.~\ref{fig:Fish_SW_d0_d1}. The red solid line is computed from Eq.~\eqref{eq:T4_all_smallJ}, including effects of the radiation transfer function, while the black dashed line shows the SW-limit result computed from Eq.~\eqref{eq:T4_all_SW}. As expected, the result including the full radiation transfer dependence is in rough agreement with the SW-limit one.} \label{fig:error_d1}
\end{figure}

Expected $1\sigma$ errors on $d_1^{\rm odd}$ are described in Fig.~\ref{fig:error_d1}. Owing to the dominance of the squeezed-limit signal, the same scaling relation as the $\tau_{\rm NL}$ case, $\Delta d_1^{\rm odd} \propto \ell_{\rm max}^{-2}$, is realized. The SW-limit result (black dashed line) shows this clearly. On the other hand, the red solid line, estimated by use of Eq.~\eqref{eq:T4_all_smallJ}, includes effects of the radiation transfer function and therefore becomes a bit bumpy. For $\ell_{\rm max} = 2000$, a minimum detectable $d_1^{\rm odd}$ is $640$, which is the same in the order of magnitude as the $\tau_{\rm NL}$ one \cite{Kogo:2006kh,Pearson:2012ba}, as expected.

\subsection{Expected uncertainties on $\rho_E^{\rm vev} / \rho_\phi$ in the $f(\phi)(F^2 + F\tilde{F})$ model}

Finally, we demonstrate the advantage of the $d_1^{\rm odd}$ measurement, by considering a concrete example: an inflationary model where the inflaton field $\phi$ couples to the U(1) gauge field $A_\mu$ via \cite{Caprini:2014mja}
\begin{eqnarray}
 S = \int d^4 x \sqrt{-g} f(\phi) \left( - \frac{1}{4} F^{\mu \nu} F_{\mu \nu} + \frac{\gamma}{4} \tilde{F}^{\mu \nu} F_{\mu \nu} \right), \label{eq:S_fFFtilde}
\end{eqnarray}
where $F_{\mu\nu} \equiv \partial_\mu A_\nu - \partial_\nu A_\mu$ and $\tilde{F}^{\mu\nu}$ are the vector kinetic term and its dual, respectively. Here we assume that the electric component of the gauge field has a vev, with a spatial fluctuation part, as ${\bf E} = {\bf E}^{\rm vev} + \delta {\bf E}$. The evolution of ${\bf E}$ and the scale dependence of the induced curvature correlators rely on the time dependence of the coupling function $f(\phi)$. We analyze almost scale-free correlators for simplicity and hence choose $f(\phi) \propto a^{-4}$ with $a(\tau)$ denoting the scale factor. Then a time-independent ${\bf E}^{\rm vev}$ is realized, and magnetic contributions are subdominant. Current CMB constraints indicate that, in this case, the energy density of the gauge field is subdominant compared with that of the inflaton field, and therefore the gauge field does not spoil stable isotropic inflationary expansion driven by the inflaton field \cite{Bartolo:2015dga}. This fact enables the perturbative treatment of effects of the interaction \eqref{eq:S_fFFtilde} on the curvature correlators. For more detailed discussions on this model, see Refs.~\cite{Caprini:2014mja,Bartolo:2015dga,Abolhasani:2015cve} and Appendix~\ref{appen:fFFtilde_zeta4}.

The resultant curvature correlators have characteristic angular dependence. Moreover, for $\gamma \neq 0$, the Chern-Simons term $\gamma F\tilde{F}$ sources the chirality of the gauge field, affecting the resultant curvature correlators.%
\footnote{See, e.g., Refs.~\cite{Dimastrogiovanni:2010sm,Shiraishi:2011ph,Bartolo:2011ee,Soda:2012zm,Maleknejad:2012fw,Bartolo:2012sd,Shiraishi:2013vja,Naruko:2014bxa} for studies on the $\gamma = 0$ regime.
} 
Reference~\cite{Bartolo:2015dga} found the expressions of the power spectrum and the (angle-averaged) bispectrum for $|\gamma| > 1$, reading
\begin{eqnarray}
 \Braket{\prod_{n=1}^2 \zeta_{{\bf k}_n} }
 &=& (2\pi)^3 \delta^{(3)}\left(\sum_{n=1}^2 {\bf k}_n \right) \nonumber \\ 
&& P_\zeta(k_1)\left[1 +  g_* ( {\hat k}_1 \cdot \hat{E}^{\rm vev} )^2 \right]
 ~, \\
 \Braket{\prod_{n=1}^3 \zeta_{{\bf k}_n} }
  &=& (2\pi)^3 \delta^{(3)}\left(\sum_{n=1}^3 {\bf k}_n \right) 
  \sum_{n} c_n P_n(\hat{k}_1 \cdot \hat{k}_2) \nonumber \\
  && P_\zeta(k_1) P_\zeta(k_2) 
  + (2~{\rm perm})
  ~, \label{eq:zeta3_def}
\end{eqnarray}
with 
\begin{eqnarray}
  g_* &\simeq& - \frac{5.4 \times 10^5}{\pi} \frac{e^{4 \pi |\gamma|}}{|\gamma|^3} \frac{0.01}{\epsilon}   \left( \frac{N}{60} \right)^2  \frac{\rho_E^{\rm vev}}{\rho_\phi} ~, \label{eq:gstar_fFFtilde} \\
  c_0 &=& - \frac{2 c_1}{3}  = 2 c_2
  \simeq \frac{0.1}{\pi}\frac{e^{4 \pi |\gamma|}}{|\gamma|^3} \frac{|g_*|}{0.01} \frac{N}{60} ~, \label{eq:cn_fFFtilde}
\end{eqnarray}
and $c_{n \geq 3} = 0$. Here, $\rho_E^{\rm vev} = E_{\rm vev}^2 / 2$ is the energy density of the gauge field vev, $\rho_\phi \simeq 3 M_p^2 H^2$ is the inflaton energy density with $M_p$ and $H$ denoting the reduced Planck mass and the Hubble parameter, respectively, $\epsilon$ is the slow-roll parameter for inflaton, $N \approx 60$ is the number of e-folds before the end of inflation at which the CMB modes leave the horizon, and $P_\zeta(k) \equiv H^2 / (4 \epsilon M_p^2 k^3)$ is the usual isotropic power spectrum due to vacuum fluctuations. In Appendix~\ref{appen:fFFtilde_zeta4} we derive the connected part of the trispectrum in the $|\gamma| > 1$ regime and find that it contains parity-odd imaginary terms as well as parity-even real ones. The imaginary part corresponds perfectly to Eq.~\eqref{eq:zeta4_dnodd_def} with 
\begin{eqnarray}
  d_0^{\rm odd} = -\frac{d_1^{\rm odd}}{ 3} \approx -\frac{0.3}{\pi^2} 
  \frac{e^{8 \pi |\gamma|}}{|\gamma|^{6}} \frac{|g_*|}{0.01} \left(\frac{N}{60}\right)^2 ~, \label{eq:dnodd_fFFtilde}
\end{eqnarray}
and $d_{n \geq 2}^{\rm odd} = 0$, while the real part is well expressed using an existing template \cite{Shiraishi:2013oqa}: 
 \begin{eqnarray} 
t_{{\bf k}_3 {\bf k}_4}^{{\bf k}_1 {\bf k}_2}({\bf K})
&=& \sum_{n} d_n^{\rm even} \left[ P_n(\hat{k}_1 \cdot \hat{k}_3) + P_n(\hat{k}_1 \cdot \hat{K}) \right. 
\nonumber \\ 
&&\left.  + P_n(\hat{k}_3 \cdot \hat{K}) \right] 
P_\zeta(k_1) P_\zeta(k_3) P_\zeta(K)
~, \label{eq:zeta4_dneven_def}
 \end{eqnarray}
 with  
\begin{eqnarray}
  d_0^{\rm even} &=& \frac{d_2^{\rm even}}{2} \approx  \frac{0.2}{\pi^2} 
  \frac{e^{8 \pi |\gamma|}}{|\gamma|^{6}} \frac{|g_*|}{0.01} \left(\frac{N}{60}\right)^2 ~, \label{eq:dneven_fFFtilde}
\end{eqnarray}
and $d_1^{\rm even} = d_{n \geq 3}^{\rm even} = 0$.

\begin{figure}[t]
    \includegraphics[width=0.5\textwidth]{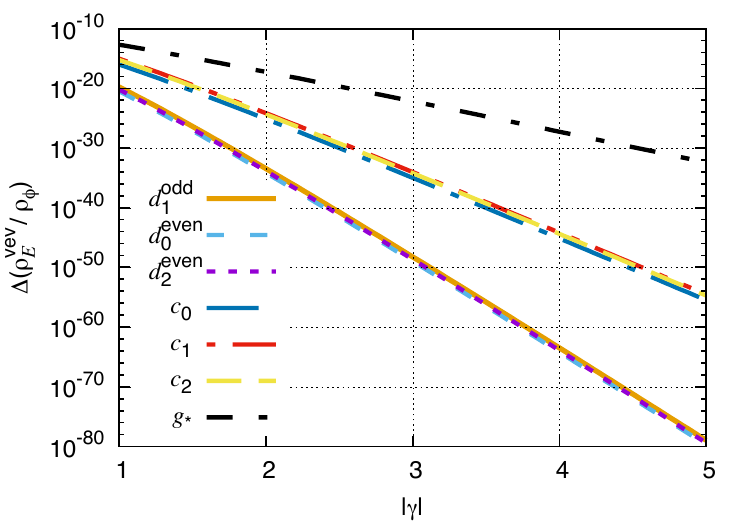}
    \caption{Expected $1\sigma$ errors on $\rho_E^{\rm vev} / \rho_\phi$ in the $f(\phi)(F^2 + F\tilde{F})$ model for $N = 60$ and $\epsilon = 0.01$, translated from $\Delta d_1^{\rm odd}$ (Fig.~\ref{fig:error_d1}), $\Delta d_{0,2}^{\rm even}$ \cite{Shiraishi:2013oqa}, $\Delta c_{0,1,2}$ \cite{Shiraishi:2013vja} and $\Delta g_*$ \cite{Ade:2015lrj} in a CVL-limit full-sky survey of the temperature anisotropies with $\ell_{\rm max} = 2000$, as a function of $|\gamma|$. Here we focus on the $ |\gamma| > 1$ regime where the approximate expressions \eqref{eq:gstar_fFFtilde}, \eqref{eq:cn_fFFtilde}, \eqref{eq:dnodd_fFFtilde} and \eqref{eq:dneven_fFFtilde} are justified. On the other hand, we do not show $|\gamma| > 5$ because such large $|\gamma|$ realizes a reheating temperature smaller than that required for successful big bang nucleosynthesis and hence is disfavored \cite{Bartolo:2015dga}.} \label{fig:error_gstar_fFFtilde_N60_lmax2000}
\end{figure}

Following the above simple relations, one can estimate an expected error $\Delta (\rho_E^{\rm vev} / \rho_\phi)$ from $\Delta d_1^{\rm odd}$, $\Delta d_{0,2}^{\rm even}$, $\Delta c_{0,1,2}$ and $\Delta g_*$. Figure~\ref{fig:error_gstar_fFFtilde_N60_lmax2000} describes $\Delta (\rho_E^{\rm vev} / \rho_\phi)$ obtained in a CVL-limit full-sky survey of the temperature anisotropies with $\ell_{\rm max} = 2000$, assuming $N = 60$ and $\epsilon = 0.01$. Concerning $\Delta d_{0,2}^{\rm even}$, $\Delta c_{0,1,2}$ and $\Delta g_*$, we adopt the results obtained in Refs.~\cite{Shiraishi:2013oqa,Shiraishi:2013vja,Ade:2015lrj}, while $\Delta d_1^{\rm odd}$ adopted here is obtained in Fig.~\ref{fig:error_d1}. In our interesting regime, $|\gamma| > 1$, the curvature correlators increase exponentially with $|\gamma|$ because of the boost of the gauge field production. This results in the exponential growth of the sensitivity to $\rho_E^{\rm vev} / \rho_\phi$ in terms of $|\gamma|$, as clearly shown in Fig.~\ref{fig:error_gstar_fFFtilde_N60_lmax2000}. In that figure we confirm the outperformance of $d_1^{\rm odd}$ as well as $d_{0,2}^{\rm even}$, compared with $c_{0,1,2}$ or $g_*$, as an observable of $\rho_E^{\rm vev} / \rho_\phi$. 

\section{Conclusions}\label{sec:conclusion}

Testing CMB parity symmetry plays an important role in the search for the primordial Universe. Parity violation in the power spectrum and bispectrum of temperature and polarization anisotropies has been widely investigated, while this paper, for the first time, studied parity violation in the CMB trispectrum. 

From the $d_n^{\rm odd}$ template of the curvature trispectrum \eqref{eq:zeta4_dnodd_def}, we derived the CMB trispectrum by means of the all-sky and flat-sky formalisms. The flat-sky expression \eqref{eq:T4_flat} contains the cross products between each $\boldsymbol{\ell}$, corresponding to those between each ${\bf k}$ in Eq.~\eqref{eq:zeta4_dnodd_def}, yielding the sign change of the CMB trispectrum under parity transformation. The nonvanishing signal obeys the parity-odd selection rule: $\ell_1 + \ell_2 + \ell_3 + \ell_4 = \text{odd}$, confirmed in the all-sky expression~\eqref{eq:T4_all}. Such a signal cannot arise from either primordial or late-time nonlinear effects in the standard parity-conserving cosmology \cite{Hu:2001fa,Kogo:2006kh}, so its detection will justify the modification or extension of the concordance framework. The $d_1^{\rm odd}$ trispectrum has a prominent signal at the squeezed configurations ($\ell_1 \sim \ell_2 \gg |\boldsymbol{\ell}_1 + \boldsymbol{\ell}_2|$ or $\ell_3 \sim \ell_4 \gg |\boldsymbol{\ell}_3 + \boldsymbol{\ell}_4|$) and hence induces a high signal-to-noise ratio comparable to the usual local $\tau_{\rm NL}$ trispectrum.

Via Fisher matrix computations, we found a minimum detectable value: $d_1^{\rm odd} = 640$, in a full-sky CVL-level survey with $\ell_{\rm max} = 2000$. With this result, we estimated a detectable value of $\rho_E^{\rm vev} / \rho_\phi$ in the $f(\phi)( F^2 + F\tilde{F})$ model. This model produces a nonvanishing power spectrum, bispectrum, parity-even trispectrum and parity-odd trispectrum, which are proportional to $\rho_E^{\rm vev} / \rho_\phi$. Comparison of these sensitivities showed the outperformance of the parity-odd trispectrum, compared with the power spectrum and bispectrum. We conclude from this that the $\ell_1 + \ell_2 + \ell_3 + \ell_4 = \text{odd}$ signal will be a promising observable of parity-violating phenomena in the inflationary epoch.

As the vector-mode or tensor-mode signal is subdominant compared with the scalar-mode one in the $f(\phi)( F^2 + F\tilde{F})$ model, we limited our analysis to the scalar sector. However, our formalism could be straightforwardly extended to the vector or tensor sector, and such an interesting issue will be addressed in future works.

\acknowledgements

I thank Nicola Bartolo, Sabino Matarrese and Marco Peloso for fruitful discussions on an anisotropic $f(\phi)( F^2 + F\tilde{F})$ model. I was supported in part by a Grant-in-Aid for JSPS Research under Grant No.~27-10917, and in part by the World Premier International Research Center Initiative (WPI Initiative), MEXT, Japan. Numerical computations were in part carried out on Cray XC30 at Center for Computational Astrophysics, National Astronomical Observatory of Japan.

\appendix
\section{Curvature trispectrum created in the $f(\phi)(F^2 + F\tilde{F})$ model}\label{appen:fFFtilde_zeta4}

In this appendix we estimate the curvature trispectrum induced from the interaction~\eqref{eq:S_fFFtilde}. Here we focus on the $|\gamma| > 1$ regime where the Chern-Simons term $\gamma F\tilde{F}$ produces the chiral gauge field effectively.

For convenience, we employ the Coulomb gauge $A_0 = \nabla \cdot {\bf A} = 0$ and the electromagnetic decomposition $E_i \equiv - \sqrt{f(\phi)} A_i' / a^2$ and $B_i = \sqrt{f(\phi)} \eta_{ijk} \partial_j A_k  / a^2$, where the prime denotes the derivative with respect to conformal time $\tau$, and $\eta_{ijk}$ is the 3D antisymmetric tensor normalized as $\eta_{123} = 1$. Let us study the case of $f(\phi) \propto a^{-4}$. This choice leads to vanishing ${\bf B}^{\rm vev}$ and time-independent ${\bf E}^{\rm vev}$ \cite{Bartolo:2015dga}. For the analysis of the fluctuation part, we move to the helicity states in Fourier space, according to
\begin{eqnarray}
  \delta E_i({\bf x},\tau) &=& \int \frac{d^3 k}{(2\pi)^{3}} \delta E_i({\bf k},\tau) e^{i {\bf k} \cdot {\bf x}} ~, \\
  \delta E_i({\bf k},\tau) &=&
  \sum_{\lambda = \pm} 
\delta E_{\bf k}^{(\lambda)}(\tau)
\epsilon_i^{(\lambda)}(\hat{k})~,
\end{eqnarray}
where $\epsilon_i^{(\lambda)}(\hat{k})$ is a divergenceless polarization vector obeying $\epsilon_i^{(\lambda)}(\hat{k}) \epsilon_i^{(\lambda')}(\hat{k}) = \delta_{\lambda, -\lambda'}$, $\hat{k}_i \epsilon_i^{(\lambda)}(\hat{k}) = 0$, $\epsilon_i^{(\lambda) *}(\hat{k}) = \epsilon_i^{(-\lambda)}(\hat{k}) = \epsilon_i^{(\lambda)}(-\hat{k})$ and $\eta_{abc} \hat{k}_a \epsilon_b^{(\lambda)}(\hat{k}) = -\lambda i \epsilon_c^{(\lambda)}(\hat{k})$. Solving the EOM of the gauge field for the $|\gamma| > 1$ regime, we notice that one of the two helicity modes increases exponentially with $|\gamma|$ due to the axial coupling $\gamma F \tilde{F}$. Without loss of generality, we can regard $\delta E_{\bf k}^{(+)}$ as the growing mode and hereinafter ignore the decaying $\delta E_{\bf k}^{(-)}$ \cite{Caprini:2014mja,Bartolo:2015dga}. In the long-wavelength regimes ($|\gamma k\tau| \ll 1$), the power spectrum can be simplified to \cite{Bartolo:2015dga}
\begin{eqnarray}
&& \Braket{\delta E_i({\bf k}_1,\tau_1) \delta E_j({\bf k}_2, \tau_2)} 
\approx (2\pi)^3 \delta^{(3)}({\bf k}_1 + {\bf k}_2)
 \nonumber \\ 
&&\qquad\qquad 
  \frac{9 H^4}{2^5 \pi} \frac{e^{4 \pi |\gamma|}}{|\gamma|^{3}} k_1^{-3}
 \epsilon_i^{(+)}(\hat{k}_1) \epsilon_j^{(+)}(\hat{k}_2) ~. \label{eq:E2_fFFtilde}
\end{eqnarray}
The magnetic mode is suppressed by $|k \tau|$ in the long-wavelength limit and hence negligible with respect to the electric mode.

Since we follow the condition that the energy density of the gauge field is subdominant with respect to the inflaton energy density, anisotropic effects on the background metric are ignorable. This enables the inflaton field $\phi$ to maintain a stable slow-roll inflation. At the same time, contributions of the gauge field to the metric fluctuation may be treated perturbatively, leading to the $N$-point curvature correlators: $\Braket{\prod_{n=1}^N \zeta_{{\bf k}_n}} = \Braket{\prod_{n=1}^N \zeta_{{\bf k}_n}}_0 + \Braket{\prod_{n=1}^N \zeta_{{\bf k}_n}}_1 + \cdots$, where the 0 mode is the contribution of the usual isotropic vacuum fluctuations \cite{Acquaviva:2002ud,Maldacena:2002vr}, and the 1 mode corresponds to the leading-order contribution due to the interaction~\eqref{eq:S_fFFtilde}. One can find the explicit expressions of the curvature power spectrum and bispectrum in Ref.~\cite{Bartolo:2015dga}. 

We now estimate the long-wavelength expression of the trispectrum generated from the interaction Hamiltonian due to Eq.~\eqref{eq:S_fFFtilde}, reading ${\cal H} =  {\cal H}_1 + {\cal H}_2$ with
\begin{eqnarray}
{\cal H}_{1}^{(\tau)} 
&\approx &   
 - \frac{4E_i^{\rm vev}}{H^4 \tau^4}  
\int \frac{d^3 p}{(2\pi)^3}  \delta E_i({\bf p}, \tau) \hat{\zeta}_{- \bf p}^{(\tau)}  
~, \\
{\cal H}_{2}^{(\tau)} 
& \approx &  
- \frac{2}{H^4 \tau^4} \int \frac{d^3 p d^3 p'}{(2\pi)^6} \nonumber \\ 
&& 
\delta E_i({\bf p}, \tau) 
\delta E_i({\bf p}', \tau) 
\hat{\zeta}_{- {\bf p} - {\bf p}'}^{(\tau)}   ~.
\end{eqnarray}
By means of the in-in formalism, the 1-mode trispectrum is written as
$\Braket{\prod_{n=1}^4 \hat{\zeta}_{{\bf k}_n}^{(\tau)}}_1 
\approx \Braket{\prod_{n=1}^4 \hat{\zeta}_{{\bf k}_n}^{(\tau)}}_1^{2211} 
+ \Braket{\prod_{n=1}^4 \hat{\zeta}_{{\bf k}_n}^{(\tau)}}_1^{2121} 
+ \Braket{\prod_{n=1}^4 \hat{\zeta}_{{\bf k}_n}^{(\tau)}}_1^{2112} 
+ \Braket{\prod_{n=1}^4 \hat{\zeta}_{{\bf k}_n}^{(\tau)}}_1^{1221} 
+ \Braket{\prod_{n=1}^4 \hat{\zeta}_{{\bf k}_n}^{(\tau)}}_1^{1212} 
+ \Braket{\prod_{n=1}^4 \hat{\zeta}_{{\bf k}_n}^{(\tau)}}_1^{1122}
$
where
\begin{eqnarray}
&& \Braket{\prod_{n=1}^4 \hat{\zeta}_{{\bf k}_n}^{(\tau)}}_1^{abcd} 
\equiv \int^\tau d\tau_1  \int^{\tau_1} d\tau_2 \int^{\tau_2} d\tau_3
\int^{\tau_3} d\tau_4 \nonumber \\ 
&& 
\Braket{ \left[\left[\left[
        \left[\prod_{n=1}^4 \hat{\zeta}_{{\bf k}_n}^{(\tau)}, {\cal H}_{a}^{(\tau_1)} \right], {\cal H}_{b}^{(\tau_2)} \right], {\cal H}_{c}^{(\tau_3)} \right], {\cal H}_{d}^{(\tau_4)} \right] 
}.
\end{eqnarray}
In the $\tau$ integrals, the small-wavelength contributions cancel each other out due to their rapid oscillating features, and hence only the long-wavelength modes survive. On such long-wavelength scales, the gauge field behaves as a classical commuting field. Owing to this fact and Wick's theorem, the expectation value in the integrand can be decomposed into the products of $\Braket{\delta E \delta E}$ and $\Braket{ \left[ \hat{\zeta}, \hat{\zeta} \right]}$. Evaluating the $\tau$ integrals with Eq.~\eqref{eq:E2_fFFtilde} and the long-wavelength expression of the commutator,
\begin{eqnarray}
 \Braket{ \left[ \hat{\zeta}_{{\bf k}_1}^{(\tau_1)}, \hat{\zeta}_{{\bf k}_2}^{(\tau_2)} \right] }
& \approx & (2\pi)^3 \delta^{(3)}({\bf k}_1 + {\bf k}_2) \nonumber \\ 
&& \left(- \frac{i H^2 }{6\epsilon M_p^2} \right)
\left( \tau_1^3 - \tau_2^3 \right)  ~, 
\end{eqnarray}
in the same manner as the bispectrum computations \cite{Bartolo:2015dga}, we obtain 
  \begin{eqnarray}
   t_{{\bf k}_3 {\bf k}_4}^{{\bf k}_1 {\bf k}_2}({\bf K}) 
    = {\cal A}_{k_3 k_4}^{k_1 k_2}(K)
  {\cal C}_{\hat{k}_3}^{\hat{k}_1}(\hat{K}, \hat{E}^{\rm vev}) \label{eq:zeta4_fFFtilde}
  ~,
   \end{eqnarray}
  where
  \begin{eqnarray}
 {\cal C}_{\hat{k}_3}^{\hat{k}_1}(\hat{K}, \hat{E}^{\rm vev})
  &\equiv& \epsilon_a^{(-)}(\hat{k}_1) \epsilon_b^{(-)}(\hat{k}_3) 
 \epsilon_a^{(-)}(\hat{K}) \epsilon_b^{(+)}(\hat{K})  \nonumber \\
 && \hat{E}_c^{\rm vev}  \hat{E}_d^{\rm vev}
 \epsilon_c^{(+)}(\hat{k}_1) \epsilon_d^{(+)}(\hat{k}_3) ~, \\
    {\cal A}_{k_3 k_4}^{k_1 k_2}(K) &=&
    \frac{9 E_{\rm vev}^2 H^{4}}{2^{12} \pi^3 \epsilon^4  M_p^8 }
  \frac{e^{12 \pi |\gamma|}}{|\gamma|^{9}}
  K^{-3}  k_1^{-3} k_3^{-3}   \nonumber \\ 
  && {\rm Min}[N_{K}, N_{k_1}, N_{k_2} ] N_{k_1}  \nonumber \\
  && {\rm Min}[N_{K}, N_{k_3}, N_{k_4} ] N_{k_3} ~, 
  \end{eqnarray}
  with $N_k \equiv - \int_{-k^{-1}}^{\tau_{\rm e}} \tau^{-1} d\tau $ denoting the number of e-folds before the end of inflation ($\tau = \tau_{\rm e}$) at which the modes with $k$ leave the horizon. Note that ${\cal A}^{k_1 k_2}_{k_3 k_4}(K) = {\cal A}_{k_1 k_2}^{k_3 k_4}(K)$ and ${\cal C}_{\hat{k}_3}^{\hat{k}_1}(\hat{K}, \hat{E}^{\rm vev} ) = {\cal C}_{\hat{k}_1}^{\hat{k}_3}(-\hat{K}, \hat{E}^{\rm vev} ) = \left[ {\cal C}_{-\hat{k}_3}^{-\hat{k}_1}(-\hat{K}, \hat{E}^{\rm vev} )\right]^*$.

  Because of the existence of $\hat{E}^{\rm vev}$ in the curvature trispectrum, the induced CMB trispectrum $\Braket{\prod_{n=1}^4 a_{\ell_n m_n}}$ breaks rotational invariance, yielding a nonvanishing signal outside the quadrilateral domain: $|\ell_1 - \ell_2| \leq J \leq \ell_1 + \ell_2$ and $|\ell_3 - \ell_4| \leq J \leq \ell_3 + \ell_4$. On the other hand, our observable $T_{\ell_3 \ell_4}^{\ell_1 \ell_2}(J)$~\eqref{eq:T4_ang_ave_def} is an angle-averaged quantity, and the anisotropic signal is prohibited. In the main text, we therefore analyze the angle-averaged form given by ${\cal C}_{\hat{k}_3}^{\hat{k}_1}(\hat{K}) = (4\pi)^{-1} \int d^2 \hat{E}^{\rm vev}
  {\cal C}_{\hat{k}_3}^{\hat{k}_1}(\hat{K}, \hat{E}^{\rm vev} )$, reading  
  \begin{eqnarray}
 24 {\cal C}_{\hat{k}_3}^{\hat{k}_1}(\hat{K})
 &=& ( \hat{k}_1 \cdot \hat{k}_3 )^2
 + ( \hat{k}_1 \cdot \hat{K} )^2  
 + ( \hat{k}_3 \cdot \hat{K} )^2 \nonumber \\ 
 && -  ( \hat{k}_1 \cdot  \hat{k}_3 )
 ( \hat{k}_1 \cdot \hat{K} )
 ( \hat{k}_3 \cdot \hat{K} ) \nonumber \\ 
 && - \hat{k}_{1} \cdot \hat{k}_3 
 - \hat{k}_1 \cdot \hat{K} 
 + \hat{k}_3 \cdot \hat{K}  \nonumber \\ 
  && - ( \hat{k}_1 \cdot \hat{K} ) ( \hat{k}_3 \cdot \hat{K} )
  - ( \hat{k}_{1} \cdot \hat{k}_3 ) ( \hat{k}_3 \cdot \hat{K} ) \nonumber \\ 
 && + ( \hat{k}_{1} \cdot \hat{k}_3 ) ( \hat{k}_1 \cdot \hat{K} )  
  \nonumber \\
  && + i \left[ \hat{k}_1 \cdot \hat{k}_3 
    + \hat{k}_1 \cdot \hat{K} 
    - \hat{k}_3 \cdot \hat{K}
    -1  \right] \nonumber \\
  &&\quad \left[ ( \hat{k}_1 \times \hat{k}_3 ) \cdot \hat{K} \right] ~. \label{eq:zeta4_C_ang_ave_fFFtilde}
  \end{eqnarray}
  It is obvious that the angular dependence in the imaginary part exactly corresponds to the $n =0$ and $1$ modes in the parity-odd isotropic template \eqref{eq:zeta4_dnodd_def}. Disregarding the logarithmic $k$ dependence in $N_k$, we obtain $d_{0,1}^{\rm odd}$ in Eq.~\eqref{eq:dnodd_fFFtilde}. Regarding the real-part contributions, one can estimate, with a simple template,
    \begin{eqnarray}
  t_{{\bf k}_3 {\bf k}_4}^{{\bf k}_1 {\bf k}_2}({\bf K}) 
  &=& {\cal A}_{k_3 k_4}^{k_1 k_2}(K) 
  \frac{0.88}{24} \nonumber \\
  && \left[ ( \hat{k}_1 \cdot \hat{k}_3 )^2 
   + ( \hat{k}_1 \cdot \hat{K} )^2  
   + ( \hat{k}_3 \cdot \hat{K} )^2 \right]. \label{eq:zeta4_dneven_temp_fFFtilde}
\end{eqnarray}
This reproduces the exact result of the Fisher matrix within an uncertainty of a few percent.%
\footnote{We compute the correlation between the CMB trispectrum from the real part of Eq.~\eqref{eq:zeta4_C_ang_ave_fFFtilde} and that from Eq.~\eqref{eq:zeta4_dneven_temp_fFFtilde} and confirm nearly $100\%$ correlation, justifying the use of Eq.~\eqref{eq:zeta4_dneven_temp_fFFtilde}.} 
Comparing this with the parity-even isotropic template \eqref{eq:zeta4_dneven_def}, we obtain $d_{0,2}^{\rm even}$ in Eq.~\eqref{eq:dneven_fFFtilde}.

\bibliography{paper}

\end{document}